\documentclass[11pt]{article}
\usepackage{graphics}
\usepackage{amssymb}
\usepackage{amsmath}
\usepackage{amscd}
\begin{document}
\title{From Quantum Action to Quantum Chaos}

\author{H. Jirari$^{a}$, H. Kr\"oger$^{a}$\thanks{invited talk of H. Kr\"oger}, G. Melkonyan$^{a}$, X.Q. Luo$^{b}$ \\
and K.J.M. Moriarty$^{c}$ \\
$^{a}$ D\'epartement de Physique, Universit\'e Laval, \\
Qu\'ebec, Qu\'ebec G1K 7P4, Canada \\
Email: hkroger@phy.ulaval.ca \\
$^{b}$ Department of Physics, Zhongshan University, 
Guangzhou 510275, China \\
Email: stslxq@zsu.edu.cn \\
$^{c}$ Department of Mathematics, Statistics and Computer Science, \\
Dalhousie University, Halifax, Nova Scotia B3H 3J5, Canada \\
Email: moriarty@cs.dal.ca \\ }

\date{}








\maketitle


\section{Introduction}
\noindent This article is about the relation between classical mechanics and quantum mechanics. The question is asked: Can quantum mechanics be formulated in such a way that it looks like some sort of classical mechanics?
Why do we ask such question in the first place? 
The answer is interesting from the point of view of interpretation of quantum mechanics. L. de Broglie \cite{Broglie} has pointed out that quantum mechanics has two faces:
The particle interpretation and the wave interpretation. 
Maybe there is a third interpretation, where quantum mechanics has the face of 
classical mechanics.
The answer is interesting also for the purpose of a proper definition, quantitative analysis and understanding of 
phenomena occuring in quantum physics, the definition of which comes from classical physics. Examples are quantum chaos and quantum instantons. 
The affirmative answer to the above question has been proposed recently in Refs.\cite{Jirari1,Jirari2}, stating that quantum transition amplitudes
can be expressed in terms of some action, called the quantum action, which has the form of the classical action but has modified parameters.

\subsection{\bf Bridges between classical mechanics and quantum mechanics} 
\noindent A general method to build bridges from classical to quantum physics is the path integral. By "bridges" we mean a relation, e.g., involving the quantum transition amplitude and the classical action. In particular, starting from the path integral, the following "bridges" have been suggested: \\

\noindent (i) {\it Sum over classical paths}. Let us consider the Q.M. transition amplitude from $x_{in}$, $t_{in}$ to $x_{fi}$, $t_{fi}$ given by 
the path integral. In certain cases this path integral can be expressed 
as a sum over classical paths only,
\begin{equation}
\label{SumClassPath}
G(x_{fi},t_{fi}; x_{in},t_{in}) = 
\sum_{ \{x_{cl}\} } Z \exp \left[ \frac{i}{\hbar} 
\left. S[{x}_{cl}] \right|_{x_{in},t_{in}}^{x_{fi},t_{fi}} \right] ,
\end{equation}
where $S[x_{cl}]$ is the classical action evaluated along the 
classical trajectory from $x_{in}$, $t_{in}$ to $x_{fi}$, $t_{fi}$. 
This is true, e.g. for the harmonic oscillator. 
Unfortunately, such relation holds only in a few exceptional cases \cite{Schulman:81}. \\

\noindent (ii) {\it Gutzwiller's trace formula}. 
Gutzwiller \cite{Gutzwiller:90} has established a relation between the 
density of states of the quantum system and a sum over 
classical periodic orbits (periodic orbit quantisation). The trace formula reads (see Ref.\cite{Blumel})
\begin{equation}
\rho(E) = \rho_{0}(E) - \frac{1}{2\pi\hbar} \mbox{Im} \sum_{p} T_{p} \sum_{n=1}^{\infty}
\frac{\exp[in (\Phi_{p}(E)/\hbar - \mu_{p} \tau/2)]}
{i \sin[n \lambda_{p}(E)/2]},
\end{equation}
where $\rho(E)=Tr[\delta(E-H)]$ denotes the density of states, 
and $\rho_0(E)$ is the average level density.
The sum runs over all primitive periodic orbits $p$, the index $n$ denotes repeated traversal of primitive periodic orbits and $T_{p}$ is the traversal time of such an orbit. 
$\Phi_{p}(E)$ is the action of the periodic orbit $p$ at energy $E$ and $\lambda_{p}(E)$ denotes a Lyapunov exponent. Here $\mu_{p}$ is a constant characteristic for the orbit $p$.
The trace formula has been applied successfully in the semi-classical regime (e.g. highly excited states of atom). Wintgen \cite{Wintgen} applied it to the diamagnetic hydrogen system and was able to extract periodic orbit information from experimental level densities. \\

\noindent (iii) The {\it effective action} 
has been introduced in quantum field theory in such a way that it gives an expectation value $<\phi>=\phi_{class}$ which corresponds to the classical trajectory and which minimizes the potential energy (effective potential). Thus one can obtain the ground state energy of the quantum system from its effective potential. The effective action $\Gamma$ \cite{Jona,Coleman} is defined by 
\begin{eqnarray}
Z[J] &=& e^{-iW[J]}
\nonumber \\
\frac{ \partial }{ \partial J(x) } W[J] &=& - <0|\phi(x)|0>_{J}
\nonumber \\
\phi_{cl}(x) &=& <0|\phi(x)|0>_{J}
\nonumber \\
\Gamma[\phi_{cl}] &=& - W[J] - \int d^{4}y ~ J(y) \phi_{cl}(y) .
\end{eqnarray}
An effective action has been also considered at finite temperature \cite{Dolan}.
Because the effective action has a mathematical structure similar to the classical action, and the quantum effects are taken into account by 
parameters different from their classical counter parts, the effective action looks like the ideal way to bridge the gap from quantum to classical physics
and eventually solve the quantum chaos and quantum instanton problem.
However, there is a catch. The effective potential and the effective action 
in quantum mechanics has been computed using perturbation theory by  
Cametti et al. \cite{Cametti}. Consider the Lagrangian
\begin{eqnarray}
L(q,\dot{q},t) &=& \frac{m}{2} \dot{q}^{2} - V(q)
\nonumber \\ 
V(q) &=& \frac{m}{2} \omega^{2} q^{2} + U(q) ,
\end{eqnarray}
and $U(q)$ is, say, a quartic potential $U(q) \sim q^{4}$.
Then the effective action is obtained in doing a loop $(\hbar)$ expansion
\begin{eqnarray}
\Gamma[q] &=& \int dt \left( - V^{eff}(q(t)) + \frac{Z(q(t))}{2} \dot{q}^{2}(t)
\right.
\nonumber \\
&+& \left. A(q(t)) \dot{q}^{4}(t) + B(q(t)) (d^{2}q/dt^{2})^{2}(t) + \cdots \right)
\nonumber \\
V^{eff} &=& \frac{1}{2}m \omega^{2} q^{2} +U(q) + \hbar V^{eff}_{1}(q) + O(\hbar^{2})
\nonumber \\
Z(q) &=& m + \hbar Z_{1}(q) + O(\hbar^{2})
\nonumber \\
A(q) &=& \hbar A_{1}(q) + O(\hbar^{2})
\nonumber \\
B(q) &=& \hbar B_{1}(q) + O(\hbar^{2}) .
\end{eqnarray}
There are higher loop corrections to the effective potential $V^{eff}$ as well as to the mass renormalisation $Z$. The most important property is the 
occurrence of higher time derivative terms. Actually, there is an infinite series of increasing order. Here comes the problem. When we want to interpret $\Gamma$ as effective action, the higher time derivatives require
more intial/boundary conditions than the classical action. 
This is a catastrophy.
In the following we will present an alternative way to construct an action 
taking into acount quantum corrections. \\

\section{Quantum Action}
\noindent We want to construct a renormalized or quantum action from transition matrix elements, which involve the time evolution. 
In quantum physics the transition amplitude from 
$x_{in}$, $t_{in}$ to $x_{fi}$, $t_{fi}$ is given by 
\begin{equation}
\label{PathIntegral}
G(x_{fi},t_{fi};x_{in},t_{in}) =
\left. \int [dx] \exp[ \frac{i}{\hbar} 
S[x] ] \right|_{x_{in},t_{in}}^{x_{fi},t_{fi}}  ,
\end{equation}
where $S$ denotes the classical action.
In Ref.\cite{Jirari1,Jirari2} we have proposed the existence of a quantum action
which satisfies the following properties: \\ 
\noindent {\it Conjecture}:
For a given classical action $S = \int dt \frac{m}{2} \dot{x}^{2} - V(x)$ 
there is a quantum action 
$\tilde{S} = \int dt \frac{\tilde{m}}{2} \dot{x}^{2} - \tilde{V}(x)$, 
which allows to express the transition amplitude by
\begin{equation}
\label{DefQuantumAction}
G(x_{fi},t_{fi}; x_{in},t_{in}) = \tilde{Z} 
\exp [ \frac{i}{\hbar} \left. \tilde{S}[\tilde{x}_{cl}] 
\right|_{x_{in},t_{in}}^{x_{fi},t_{fi}} ] .
\end{equation}
Here $\tilde{x}_ {cl}$ denotes the classical path, such that the action $\tilde{S}(\tilde{x}_{cl})$ 
is minimal (we exclude the occurrence of conjugate points or caustics). 
$\tilde{Z}$ denotes the normalisation factor corresponding to $\tilde{S}$. 
Eq.(\ref{DefQuantumAction}) is valid with 
the {\em same} action $\tilde{S}$ for all sets of 
boundary positions $x_{fi}$, $x_{in}$ for a given time interval $T=t_{fi}-t_{in}$. 
The parameters of the quantum action depend on the time $T$. The 
quantum action converges to a non-trivial limit when $T \to \infty$. 
Any dependence on $x_{fi}, x_{in}$ enters via the trajectory 
$\tilde{x}_ {cl}$. $\tilde{Z}$ depends on the action parameters and $T$, 
but not on $x_{fi}, x_{in}$. \\
 
\noindent One may ask: What is the difference between effective and quantum action?
Conceptually, effective action and 
quantum action look quite similar.
However, its technical definition is different and also its physical content. The effective action requires $<\phi>=\phi_{cl}$, while the quantum action does not. The effective action corresponds to infinite time 
and allows to obtain the ground state energy, but the quantum action 
is defined for arbitrary finite time $T$.  In Euclidean formulation, the inverse time corresponds to temperature. Thus the quantum action allows to describe quantum physics at finite temperature including excited states (see below). However, the effective action can be defined also at finite temperature \cite{Dolan}.
The effective action can be computed analytically by perturbation theory (loop expansion). However, this series is not convergent. 
Practically, it can be used only for some small number of loops and small values of the perturbation parameter.
The quantum action can be computed non-perturbatively for all values of the coupling parameter. 
The effective action has the defect of generating higher order time derivatives.
The quantum action is postulated to be free of higher time derivative terms. 
To construct the quantum action being sensitive to excited states, one needs transition matrix elements beyond the vacuum sector. We have chosen to use position states in Q.M. In Q.F.T. this corresponds to Bargman states.

\subsection{Construction of quantum action}
\noindent Suppose the classical action is given by
\begin{equation}
S = \int_{0}^{T} dt \frac{m}{2} \dot{x}^{2} - v_{4} x^{4}(t).
\end{equation}
Then we make an ansatz for the quantum action
\begin{equation}
\tilde{S} = \int_{0}^{T} dt \frac{\tilde{m}}{2} \dot{x}^{2} 
- \left\{ \tilde{v}_0 + \tilde{v}_{1} x(t) + \cdots + \tilde{v}_{N} x^{N}(t) \right\}.
\end{equation}
Then $\tilde{m}$, $\tilde{v}_0$, \dots, $\tilde{v}_{N}$ are the renormalized parameters which take into account the quantum corrections. 
Their values are determined by making a global best fit to a number of transition amplitudes $G(x_j,T;x_i,0)$ (which satisfies Eq.\ref{DefQuantumAction}), where $x_i$, $x_j$ haven been taken from a set of points $\{x_1,\cdots,x_J\}$ and those points have been chosen to cover
some interval $[-a,+a]$.  
More details are given in Refs.\cite{Jirari1,Jirari2}.
As an example, 
the parameters of the quantum action 
corresponding to the double well potential (action $S = \int_{0}^{T} dt 
\frac{1}{2} m \dot{x}^{2} - \{v_{0} + v_{2} x^{2} + v_{4} x^{4} \}$,  
$v_{0} = \frac{1}{2}$, $v_{2} = -1$, $v_{4} = \frac{1}{2}$) as function of $T$ is shown in Fig.[1]. One observes that the parameters of the quantum action vary with the transition time $T$.
For small $T$ (limit $T \to 0$) the parameters of the quantum action are consistent with those of the classical action. For sufficiently large time $T$, the parameters of the quantum action tend to converge asymptotically.

\subsection{Quantum action at finite temperature}
\noindent First we make a Wick rotation to imaginary time. 
The purpose is, first to make the path integral well defined (Wiener measure)
allowing to apply Monte Carlo methods for its numerical computation. Secondly, the instanton is defined in imaginary time.
One effect of this transformation is that it changes a relative sign between 
the kinetic term and the potential term of the action. 
Thus in the following we work with imaginary time (Euclidean) actions and 
Green's functions.
Let us see how the quantum action is related to finite temperature physics. 
According to the laws of quantum mechanics and thermodymical equilibrium,
the expectation value of some observable $O$, like e.g. average energy
is given by
\begin{eqnarray}
<O> &=& \frac{ Tr\left[ O ~ \exp[ - \beta H] \right] }
{ Tr\left[ \exp[ - \beta H] \right] }
\nonumber \\
&=& \frac{ \int_{-\infty}^{+\infty} dx \int_{-\infty}^{+\infty} dy
<x|O|y><y|\exp[ - \beta H ]|x> }
{ \int_{-\infty}^{+\infty} dx <x|\exp[ - \beta H ]|x> } ,
\end{eqnarray}
where $\beta$ is related to the temperature $\tau$ by 
$\beta = 1/(k_{B} {\tau})$.
On the other hand the (Euclidean) transition amplitude is given by
\begin{equation}
G(x_{fi},T;x_{in},0) = <x_{fi}| \exp[ - H T/\hbar ]|x_{in}>
\end{equation}
Thus from the definition of the quantum action, Eq.(\ref{DefQuantumAction}),
one obtains
\begin{equation}
<O> = \frac{ \int_{-\infty}^{+\infty} dx \int_{-\infty}^{+\infty} dy
<x|O|y> \exp[-\tilde{S}_{\beta}|_{x,0}^{y,\beta}] }
{ \int_{-\infty}^{+\infty} dx \exp[ - \tilde{S}_{\beta}|_{x,0}^{x,\beta}] } ,
\end{equation}
if we identify 
\begin{equation}
\beta = \frac{1}{k_{B} {\tau}} = T/\hbar .
\end{equation}
As a result, the quantum action $\tilde{S}_{\beta}$ computed from transition time $T$, 
describes equilibrium thermodynamics at $\beta = T/\hbar$, i.e. temperature $\tau = 1/(k_{B} \beta)$. \\

\noindent In the case of the double well potential we have found that parameters of the quantum action vary as function of $T$ (Fig.[1]). Translating this behavior into temperature, it means that the parameters of the quantum action are temperature dependent (or $\beta$-dependent). 
In particular, we can interpret the behavior for small $T$ as follows.
$T=0$ means temperature $\tau = \infty$. The quantum action at infinite temperature coincides with the classical action. On the other hand, 
the limit $T \to \infty$ corresponds to temperature $\tau \to 0$. The quantum behavior is dominated by the ground state (Feynman-Kac formula).

\section{Quantum Instantons}
\noindent Quantum mechanics, which describes physics at atomic length scales can not be understood by the laws of classical physics valid at macroscopic length scales.
Examples are: Heisenberg's uncertainty principle, quantum tunneling, Schr\"odinger's cat paradox, entangled states, Einstein-Rosen-Podolski paradox, quantum cryptology, quantum computing etc. On the other hand, in modern physics there are notions which have proven to be quite useful and which have their origin in classical physics.
For example consider instantons. Instantons play a role in quantum chromodynamics (QCD), the standard model of strong interactions. They may be important for the mechanism of confinement of quarks. Presumably they play an important role in nuclear matter at high temperature and density, where a phase transition from the hadronic phase to the quark-gluon plasma has been predicted. Even a richer phase structure may exist \cite{Shuryak}. Furthermore, in the inflationary scenario of the early universe, instantons are important. For a review see Ref.\cite{Linde}. 
During inflation, quantum fluctuations of the primordial field expand exponentially and eventually end up as a classical field. The fluctuations are of the size of the horizon \cite{Starobinsky}. The classical fluctuations eventually lead to galaxy formation \cite{Khlopov}. \\

\noindent In quantum physics, an instanton solution is conventionally defined as the saddle point approximation of the (Euclidean) path integral. 
However, there is a problem with the proper definition of instantons in quantum physics: Let us consider a 1-D system in quantum mechanics with a particle of mass $m$ moving in a potential $V(x)=A(x^{2}- a^{2})^{2}$. This potential has two minima at $x=\pm a$. The instanton $x_{inst}(t)$ is the solution of the classical equation of motion in imaginary time, with boundary conditions such that the particle starts at $x(t=-\infty)=-a$, $\dot{x}(t=-\infty)=0$ and arrives at $x(t=+\infty)=+a$, $\dot{x}(t=+\infty)=0$. The problem again is that quantum mechanics does not allow to specify both, position and momentum with zero uncertainty. \\

\noindent In Ref.\cite{Jirari2} we have suggested to define a quantum instanton solution via the quantum action. This means to compute non-perturbatively the quantum action $\tilde{S}$ (in imaginary time) and analyze if the corresponding quantum potential $\tilde{V}$ has multiple degenerate minima (degenerate vacua).
Then the quantum instanton is defined as the classical solution $\tilde{x}_{class}$ between those minima (there is no problem with boundary conditions).
Such quantum instanton solutions have been computed in quantum mechanics for the 1-D double well potential in Ref.\cite{Jirari2}. 
The quantum instantons were found to be "softer" than the classical instantons (potential minima were closer and the potential barrier was lower).

\section{Quantum Chaos}
\noindent Classical deterministic chaos has been observed in a huge number of phenomena in macroscopic i.e. classical physics. But chaotic phenomena were also found in systems ruled by quantum mechanics. 
For example, the hydrogen atom in a strong magnetic field 
shows strong irregularities in its spectrum \cite{Friedrich}.
Irregular patterns have been found in the wave functions of the quantum mechanical model of the stadium billard \cite{McDonald}.
Billard like boundary conditions have been realized experimentally 
in mesoscopic quantum systems, like quantum dots and quantum corrals, formed by atoms in semi-conductors \cite{Stockmann}. \\

\noindent So what is the problem with chaos in quantum physics?
It has to do with its proper definition. The underlying reason is due to the dynamical group of time evolution. In classical mechanics time evolution of a system can be viewed as an infinite sequence of infinitesimal canonical transformations. The corresponding dynamical group is the symplectic group. In quantum mechanics, a system governed by a time independent Hamiltonian, follows the time evolution of the unitary group. This difference has simple but drastic consequences: In classical physics, chaos is characterized, e.g. by Lyapunov exponents or Poincar\'e sections. This is based on identifying trajectories in phase space (position and conjugate momentum).
In quantum mechanics, Heisenberg's uncertainty relation $\Delta x \Delta p \ge 
\hbar/2$ does not allow to specify a point in phase space with zero error!
Consequently, the apparatus of classical chaos theory can not be simply taken over to quantum physics. \\

\noindent Due to this problem, workers in quantum chaos have tried to characterize such systems in different ways, alternative to those of classical chaos. One successful route has been to characterize the spectral density of quantum system with chaotic classical counterpart by Poisson versus Wigner distributions. There is a conjecture by Bohigas et al. \cite{Bohigas}, which says that the signature of a classical chaotic system is a the spectral density following a Wigner distribution. \\

\subsection{Quantum chaos in 2 dimensions}
\noindent As the problem with a proper definition of quantum chaos 
has the same root as the problem with quantum instantons, we suggest also to apply the same strategy of solution, i.e. define quantum chaos via the quantum action. Then the quantum action $\tilde{S}$ incorporates the effects of quantum physics, but has mathematically the structure of a classical action. 
The apparatus of classical chaos theory, like Lyapunov exponents, Poincar\'e sections etc. can be applied to the quantum action $\tilde{S}$. \\

\noindent As is well known 1-dimensional conservative systems with a time-independent Hamiltonian are integrable and do not produce classical chaos.
An interesting candidate to consider is the K-system, corresponding to the potential $V=x^{2}y^{2}$. This decribes a 2-D Hamiltonian system, being almost globally chaotic, having small islands of stability \cite{KSystem}. However, from the numerical point of view more convient, but also showing classical chaos, is the following related system, investigated by Pullem and Edmonds \cite{Pullen}, It is defined by the classical action
\begin{equation}
S = \int_{0}^{T} dt ~ \frac{1}{2} m (\dot{x}^{2} + \dot{y}^{2}) 
- ( v_{2+2}(x^{2} + y^{2}) + v_{22} x^{2}y^{2} ).
\end{equation}
As parameters of the classical action we use $m=1$, $v_{2+2}=0.5$, $v_{22}=0.05$ and the convention $\hbar=k_{B}=1$.
The Poincar\'e sections corresponding to energies $E=10, 20, 50$ 
are shown in Figs.[2a,3a,4a]. \\

\noindent For the corresponding quantum action, we make the following ansatz,
which is compatible with time-reversal symmetry, parity conservation and symmetry under exchange $x \leftrightarrow y$,
\begin{eqnarray}
\tilde{S} &=& \int_{0}^{T} dt ~ 
\frac{1}{2} \tilde{m}_{2+2} (\dot{x}^{2} + \dot{y}^{2}) 
+ \frac{1}{2} \tilde{m}_{11} \dot{x} \dot{y}
\nonumber \\
&-& \left\{ \tilde{v}_{0} 
+ \tilde{v}_{11} x y 
+ \tilde{v}_{2+2} (x^{2} + y^{2}) 
+ \tilde{v}_{22} x^{2}y^{2} 
+ \tilde{v}_{1+3} (x y^{3} + x^{3} y)
\right.
\nonumber \\
&+& \left. \tilde{v}_{4+4} (x^{4} + y^{4}) \right\} .
\end{eqnarray}
We have determined numerically the parameters of the quantum action 
for transition time $T=0.5$, corresponding to temperature $\tau=2$, and find  
\begin{eqnarray}
\tilde{m}_{2+2} &=& 0.9998(2)
\nonumber \\
\tilde{m}_{11} &=& 0.0000(3) 
\nonumber \\ 
\tilde{v}_{0} &=& 1.1875(32)
\nonumber \\ 
\tilde{v}_{11} &=& 0.0105(31)
\nonumber \\ 
\tilde{v}_{2+2} &=& 0.5098(63)
\nonumber \\ 
\tilde{v}_{22} &=& 0.0523(15)
\nonumber \\ 
\tilde{v}_{1+3} &=& 0.0016(12)
\nonumber \\ 
\tilde{v}_{4+4} &=& 0.0017(30) .
\end{eqnarray} 
The data are consistent with vanishing parameters $\tilde{m}_{11}$, 
$\tilde{v}_{1+3}$ and $\tilde{v}_{4+4}$. The quantum action slightly modifies the parameters $\tilde{v}_{2+2}$ and the parameter $\tilde{v}_{22}$. 
We computed the Poincar\'{e} sections for the quantum action at temperature $\tau=2$, corresponding to energies $E=10, 20, 50$. They are shown in   Figs.[2b,3b,4b]. One observes that the quantum system also displays chaos, and the Poincar\'{e} sections are slightly different from those of the classical action. One should note that the classical action at $T=0$  is equivalent to a quantum action at temperature $\tau=\infty$.

\section{Discussion}
\noindent We have discussed the use of the quantum action, which can be considered as a renormalized classical action at finite temperature. We found that the quantum action 
solves the problem of proper definitions of quantum instantons and quantum chaos.
As an example, we have considered harmonic oscillators with a weak anharmonic coupling ($V_{coupl} \sim x^2 y^2$) and computed the quantum action at temperature $\tau=2$. We compared Poincar\'{e} sections at temperature $\tau=\infty$ and $\tau=2$ and found that both display chaos. \\

\noindent {\bf Acknowledgements} \\ 
H.K. and K.M. are grateful for support by NSERC Canada. 
X.Q.L. has been supported by NSF for Distinguished Young Scientists of China, by Guangdong Provincial NSF and by the Ministry of Education of China.
H.K. is grateful for discussions with L.S. Schulman.

\vspace{1.0cm}

\noindent Figure caption \\
\noindent Fig.[1] Quantum action parameters corresponding to classical double well potential $V(x) = \frac{1}{2}(x^{2}-1)^{2}$.
Quantum action parameters $\tilde{m}$ (full diamond, full dot), $\tilde{v}_{0}$ (open diamond, open dot), $\tilde{v}_{2}$ (full triagles), $\tilde{v}_{4}$ (open triangles) versus transition time $T$ (inverse temperature). \\

\noindent Fig.[2a] Classical Poincar\'e sections corresponding to the classical potential $V(x,y)=\frac{1}{2}(x^{2} + y^{2}) + 0.05 x^{2} y^{2}$. 
Energy $E=10$. \\

\noindent Fig.[2b] Poincar\'e sections from the quantum action corresponding to the classical potential $V(x,y)=\frac{1}{2}(x^{2} + y^{2}) + 0.05 x^{2} y^{2}$. 
Energy $E=10$. \\

\noindent Fig.[3a] Like Fig.[2a], but energy $E=20$. \\

\noindent Fig.[3b] Like Fig.[2b], but energy $E=20$. \\

\noindent Fig.[4a] Like Fig.[2a], but energy $E=50$. \\

\noindent Fig.[4b] Like Fig.[2b], but energy $E=50$. \\

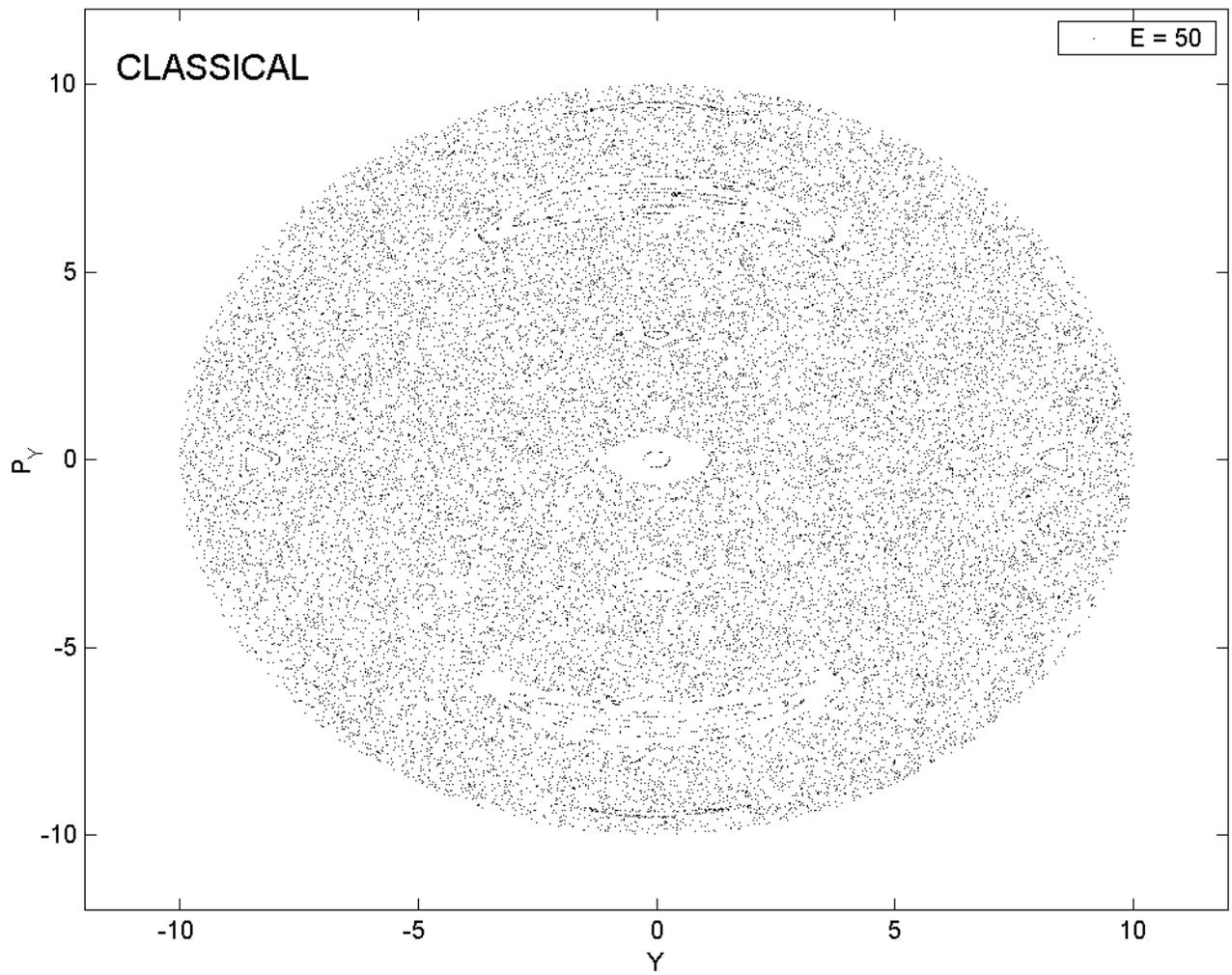
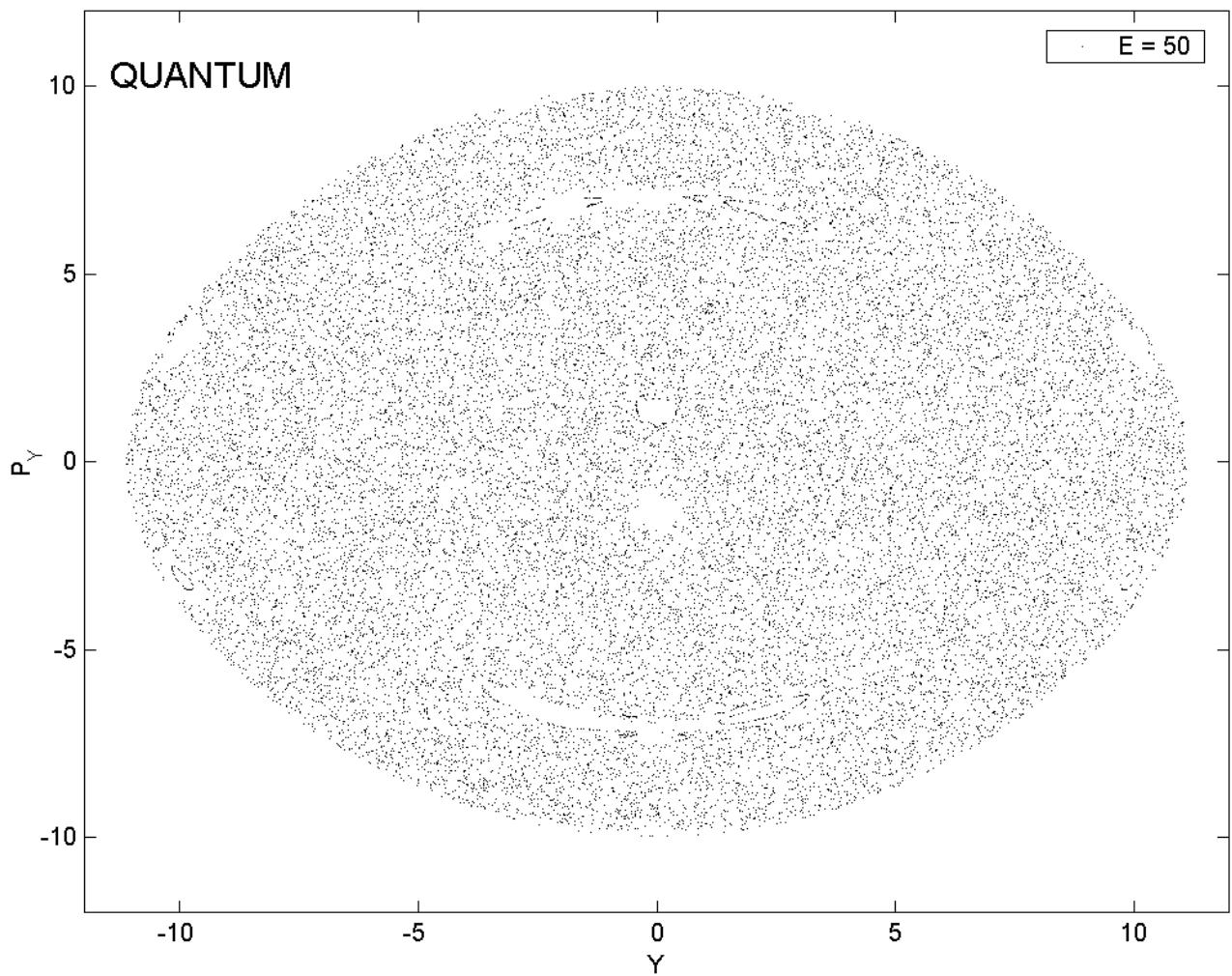
\end{document}